\documentstyle[12pt,aaspp]{article}	
\input{psfig.sty}

\def\ni{\noindent}

\def\about{$\sim$}
\def\arcsec{$\,^{\prime\prime}$}
\def\arcmin{$\,^\prime$~}

\def\erg/cm2sec{ergs~cm$^{-2}$~s$^{-1}$}  
\def\ergcm2{ergs~cm$^{-2}$}  
  
\def\mdot{$\dot{m}$~}  
\def\X{$\times$~}

\def\FxFv{{F$_x$/{F$_v$}}~}
\def\Lx{L$_x$~}
\def\Lxh{L$_x$(2-10keV)~}
\def\pc3{pc$^{-3}$~}
\def\rc{r$_c$~}

\def\apj{{\it Astrophys. J.}}
\def\aap{{\it Astron. Astrophys.}}
\def\mn{{\it Mon. Not. R. Astron. Soc.}}
\def\pasp{{\it Publ. Astron. Soc. Pacific}}

\def\avcolor{$\overline{Xcolor}$~}
\def\sigcolor{$\sigma_{xcolor}$~}

\newcommand{\lsim }{{\lower0.8ex\hbox{$\buildrel <\over\sim$}}}
\newcommand{\gsim }{{\lower0.8ex\hbox{$\buildrel >\over\sim$}}}

\newcommand{\Msun}{\ifmmode {M_{\odot}}\else${M_{\odot}}$\fi~}
\newcommand{\Rsun}{\ifmmode {R_{\odot}}\else${R_{\odot}}$\fi}
\newcommand{\Lsun}{\ifmmode {L_{\odot}}\else${L_{\odot}}$\fi}
\newcommand{\mv}{\ifmmode {m_{V}}\else${m_{V}}$\fi}
\newcommand{\Mv}{\ifmmode {M_{V}}\else${M_{V}}$\fi}
\newcommand{\lopt}{\ifmmode L_{opt} \else $~L_{opt}$\fi}
\newcommand{\loglopt}{\ifmmode{\rm log}~L_{opt} \else log$~L_{opt}$\fi}
\newcommand{\lx}{\ifmmode L_x \else $~L_x$\fi}
\newcommand{\loglx}{\ifmmode{\rm log}~L_x \else log$~L_x$\fi}
\newcommand{\cmsq}{\ifmmode{\rm ~cm^{-2}} \else cm$^{-2}$\fi}
\newcommand{\nh}{\ifmmode{\rm N_{H}} \else N$_{H}$\fi}
\newcommand{\fcgs}{\ifmmode {\rm erg~cm}^{-2}~{\rm s}^{-1}\else
erg~cm$^{-2}$~s$^{-1}$\fi} 
\newcommand{\lcgs}{\ifmmode erg~~s^{-1}\else erg~s$^{-1}$\fi}

\begin{document}

\vspace*{-4.cm}

\small
\begin{center}
Published online in {\it Science Express} on May 17, 2001\\
reference: 10.1126/science.1061135 \\
(to be published in {\it Science} (print version) in June, 2001.
\end{center}
\normalsize


\Large
\begin{center}
 {\bf High Resolution X-ray Imaging of a Globular Cluster Core: Compact 
Binaries in 47Tuc}
\end{center}
\medskip

\large
\begin{center}
{\bf Jonathan E. Grindlay\footnote{To whom correspondence should be 
addressed. E-mail: josh@cfa.harvard.edu}, Craig Heinke, Peter D. 
Edmonds, and Stephen S. Murray}
\end{center}
\affil{Harvard-Smithsonian Center for Astrophysics, 
60 Garden Street, Cambridge, MA  02138}
\medskip

\normalsize
\noindent
{\bf 
We have obtained high resolution (\lsim1\arcsec) deep 
x-ray images of the globular cluster 47Tucanae (NGC 104) 
with the Chandra X-ray Observatory to study the 
population of compact binaries in the high stellar density core. 
A 70 kilosec exposure of the cluster reveals a 
centrally concentrated population of faint (\Lx \about10$^{30-33}$ \lcgs)  
x-ray sources, with at least 
108 located within the central 2\arcmin \X 2.5\arcmin and 
\gsim half with \Lx \lsim10$^{30.5}$ \lcgs. 
All 15 millisecond pulsars (MSPs) recently located precisely by radio 
observations are identified, though two are unresolved by Chandra. 
The x-ray spectral and temporal characteristics,  
as well as initial optical identifications with 
the Hubble Space Telescope,   
suggest that \gsim50\% are millisecond pulsars, \about30\% are accreting 
white dwarfs, \about15\% are 
main sequence binaries in flare outbursts and only 2 to 3 are quiescent 
low mass x-ray binaries containing neutron stars, 
the conventional progenitors of MSPs. 
An approximate upper limit of \about470\Msun 
for the mass of an accreting central black hole in the cluster is derived. 
These observations provide the first x-ray ``color-magnitude" diagram 
for a globular cluster and census of its compact object and binary 
population.
}

\medskip

As the oldest stellar systems in the Milky Way Galaxy, globular clusters 
are laboratories for studies of stellar and dynamical evolution. 
The advantages of measurable ages and distances, both 
of which can be determined currently to \about10\%, make globulars 
an especially attractive site for study of the evolution of 
stellar populations. The high stellar densities found in the 
cores of many globulars, with central values as high as 
10$^6$\Msun\pc3 (1), implies that 
stellar interactions or near encounters can be relatively common -- 
especially between stars and binary star systems. The role 
of binaries in cluster cores is dramatic: as a source of localized 
gravitational binding energy, they act as a dynamical heat 
source in the cluster core, stabilizing it against further collapse  
into a central black hole (2). 

Compact binary stars, in which one member is either a white dwarf (WD) or 
a neutron star (NS), are markers of the extremes of stellar and binary 
evolution. Together with short-period binaries containing main sequence 
stars, compact binaries interact with stars in globular cluster cores 
and affect the stellar populations directly. Stellar evolution 
within compact binaries containing either a WD or NS 
leads to mass transfer, and accretion onto the compact object makes them 
(and their progeny, MSPs) visible as persistent x-ray sources. X-ray flare 
emission, and fainter continuous emission, can also be 
detected from chromospherically active main sequence (M-S) binaries,
known as BY Draconis systems. 
Thus x-ray observations are a particularly efficient way to study 
the compact binary population and the WD vs. NS population in globulars.

A complete sample of luminous x-ray sources (\Lxh \gsim10$^{36}$ \lcgs) 
has been discovered  in globular clusters in 
the Galaxy: one bright source is 
detected (though nearly 50\% are transient) in each of 12 clusters,  
most with high central stellar density and 
metallicity (3). Type I x-ray bursts (4) have been detected from each, 
indicating all are NSs accreting from lower mass binary companions. This 
was also suggested by the \about3\arcsec-accuracy 
x-ray positions measured for 6 of them with 
the Einstein x-ray Observatory which led to 
subsequent optical identifications for several 
and which yielded radial offsets 
consistent with their being \about2\Msun objects (5). These 12 low mass x-ray 
binaries (LMXBs) in the system of \about150 globulars in the Galaxy are, 
as originally suspected (6), \about200\X more common (per unit mass) 
in globulars than in the Galaxy as a whole. Thus 2-body stellar (tidal) 
interactions between a NS and M-S star, and the more probable exchange 
interactions of a NS with a pre-existing binary of two M-S stars, 
enhances LMXB production in globulars. The likely descendants of LMXBs, 
MSPs, in which the NS 
has been spun up to millisecond spin periods by accretion and becomes visible 
as a radio pulsar after the accretion phase ends (7), are 
also favored in globulars. The relative LMXB vs. MSP populations 
in a globular can thus constrain the formation epoch of each and 
whether alternative production channels are required for MSPs, such as 
accretion induced collapse (AIC) of massive WDs (to NSs) which could 
directly produce MSPs without a progenitor LMXB phase (7). 
In the relatively massive and high 
central density cluster 47Tuc (NGC 104), at least 20 MSPs are now 
detected which imply a total population of \about100--200 in this 
cluster alone (8). 

We obtained a 70 ksec exposure of 47Tuc on UT16.31 - 17.22 March, 2000, 
with the Chandra X-ray Observatory and Advanced 
CCD Imaging Spectrometer-Imager 
(ACIS-I) at its focus (9). Our primary goal was to study the 
population of low luminosity x-ray sources (\Lx \lsim 10$^{33-34}$ \lcgs) 
discovered in this and other globulars (10) and originally suggested to be 
primarily accreting WDs and thus cataclysmic variables (CVs), together 
with a smaller population of the typically transient NS systems (LMXBs) in 
quiescence (quiescent LMXBs). Alternatively, the quiescent LMXB population may 
dominate the faint x-ray source population 
in globulars (11), and be required to maintain the 
large MSP population if they are derived continuously from LMXBs (7). 
Thus measurement and identification of a large sample of low luminosity 
x-ray sources in 47 Tuc, with its rich MSP population, can constrain 
the formation and evolution of MSPs as well 
as the relative populations of cluster WDs vs. NSs, with implications 
for the cluster initial mass function.  
We present results from a 2.0\arcmin \X 2.5\arcmin field 
centered on the cluster core and chosen to include all of the 15 MSPs 
with precise positions (12). This analysis for an 
inscribed radius of 1\arcmin = \about2.6r$_c$, where 
r$_c$ is the cluster core radius (13), should include 
most compact binaries given their expected mass segregation (2). Results 
for the full 16\arcmin \X 16\arcmin ACIS-I field will 
be presented subsequently. 
\medskip

\ni
{\bf Overview of Source Distributions}

Previous x-ray studies of 47Tuc with the ROSAT telescope and HRI  
detector (14) revealed a population of 9 sources with 
x-ray luminosity \Lx \about 
3-100 \X 10$^{31}$ \lcgs within \about1\arcmin of the 
cluster center as well as underlying unresolved emission 
with total luminosity \Lx \about 4 \X 10$^{32}$ \lcgs. The 
factor of \gsim10 improvements in angular and spectral resolution (to 
\about0.5\arcsec~ and \about120eV, respectively, with ACIS) achieved 
by Chandra over ROSAT enable detection of sources \about10-100\X 
fainter. 
The spectral resolution enables comparison with x-ray spectra of known
classes of source, while the exquisite positional resolution 
permits much more detailed searches for optical counterparts with
high resolution HST images. Our deep exposure reached a sensitivity 
limit of \Lx \about 6 \X 10$^{29}$ \lcgs for a threshold source 
with 3 counts (vs. \about0.5 counts total background per source) 
detected at 90\% confidence by the Chandra analysis tool wavdetect (15) 
and assuming a soft thermal bremsstrahlung spectrum with 
temperature kT = 1 keV (see below) for comparison with the ROSAT results (14). 
The ``true" color image of the cluster core as imaged by Chandra (Fig. 1) 
shows the range of source brightness and 
spectral color by combining counts detected in soft (0.5 - 1.2 keV; red), 
intermediate (1.2 - 2 keV; green) and hard (2 - 6 keV; blue) bands. Wavdetect 
finds 103 sources with total counts (in a 0.5 - 4.5 keV primary 
detection band) from 3 (detection threshold) to 5332 over the total 70 ksec 
exposure. Five additional sources are found at the positions of MSPs 
Q, L, T, M and C: Q, L, and T were missed by wavdetect due 
to crowding, while M and C are only 3 and 1 count ``detections", 
though precisely at the MSP positions, 
due to low exposure (\about30\%) on the 
detector gaps. Using the recently derived distance (5.0kpc) 
and absorption for 47Tuc (16), 
we derive the luminosity distribution for the sources (Fig. 2). 
The complete list of 108 source positions and counts 
detected in several bands, as well as source  
identifications currently derived (some discussed below), 
are given in the Supplementary Table (available at {\it Science} On Line).

The x-ray luminosity function increases 
at low values, with more than half the sources with \Lx \lsim 
3 \X 10$^{30}$ \lcgs. The 23 sources in the lowest \Lx bin 
(2.5 - 7.5 cts; \Lx\about10$^{30}$ \lcgs) of the linear distribution 
form an incomplete set. From visual examination of the image as 
well as a \about500 count excess of total counts over 
detected source counts  
in the central core region (r \lsim1\rc), 
we estimate an additional \gsim100 sources with 
\Lx \lsim10$^{30}$ \lcgs 
are missed with wavdetect due to crowding. Both  
a statistics-limited overlapping source 
detection algorithm (17) and a deeper ACIS-S exposure 
(with better low energy sensitivity) are needed 
to constrain the number and types of these faintest core sources.

The sources appear to be of (at least) 4 types as marked with 
different symbols in Fig. 1 and discussed below. The distribution 
of source types can also be seen in the x-ray ``color magnitude" 
 diagram (Fig. 3), which is possible to derive for the first 
time for a significant population of  x-ray sources in a single 
globular cluster. 

\medskip

\ni
{\bf Source Identifications}

Our analysis region of the cluster core was chosen to include the 
15 MSPs with known positions (12) because 10 MSPs (including one 
in the globular cluster M28) have been detected in x-rays (18) and 
detection of even some in 47Tuc would enable a study of a 
population of cluster MSPs and their relation to both quiescent LMXBs and CVs. 
All 14 of the resolvable MSPs are detected (MSPs G and I, possibly a 
bound triple, have only 0.12\arcsec~ separation (12)), 
with exposure-corrected counts ranging from 3-27 and thus \Lx \about 
10$^{29.8-30.8}$ \lcgs. Source identifications are possible, since 
even the faintest (3ct; \gsim2$\sigma$ in wavdetect) Chandra sources have 
positions with \about0.2\arcsec~ uncertainties (1$\sigma$) due to 
the \about0.8\arcsec~ Chandra image diameter (near the telescope 
axis) and 0.5\arcsec~ ACIS pixel size.  The 
precise (\about0.001\arcsec) radio timing positions of the 
MSPs permit an astrometric solution for the Chandra 
source positions. Using 7 of the brighter 
or best resolved Chandra/MSP source candidates, the rms 
deviation between the Chandra and precise MSP 
positions is only 0.11\arcsec. Given a similarly precise Chandra-HST 
astrometric solution from our optically identified sources (see 
below), this will permit optical searches for the MSP counterparts.  
Because the 14 resolved Chandra MSP candidates are all 
located near the center of the circles which mark the 
precise MSP positions (Fig. 1), the identifications are very likely. 
Although the probability that one of the estimated 
\about100 unresolved sources 
in the 23\arcsec~ radius core is within 0.5\arcsec of any MSP 
position is 0.05, this is applicable to only MSPs L and T (the only 
MSPs in the core not 
detected automatically with wavdetect, and thus with positional 
uncertainties \about1 pixel = 0.5\arcsec~ instead of wavdetect 
centroid values of typically \lsim0.2\arcsec). 
The ACIS time resolution (nominal 3.2sec) does not permit 
analysis for pulsations, and the limited 
counts do not permit spectral analysis other than the 
determination of hardness ratios. 
The relatively narrow x-ray luminosity range implies  
a  steep x-ray luminosity function, with typical \Lx similar 
to the field MSPs detected outside of clusters (18), 
though the predominantly ``red" x-ray colors (see below) 
are in contrast to some of the 
field MSPs with hard spectral components and the one in M28 with \Lx 
\about10$^{32-33}$\lcgs (18). Full details of the 47 Tuc MSP x-ray 
source characteristics will be presented separately (19). 

Three other classes of sources are suggested by their x-ray spectra 
and temporal characteristics (Fig. 1). X5 and X7 are probable 
quiescent LMXBs, 13 sources are candidate CVs, and 6 
sources appear to be main sequence (M-S) binaries detected in flare 
outbursts due to enhanced chromospheric activity. Each class is considered 
in turn to then enable constraints on the much larger unidentified 
population. 

X5 and X7 are detected in the 0.5-4.5keV 
medium-energy band with 4435 and 5332 counts (hereafter medcts) and 
have spectra dominated by soft blackbody  
components with kT= 0.31 and 0.29 
keV, respectively. Their apparently constant 
black body luminosities (log \Lx = 32.8; 
comparable to that detected with ROSAT (14)) and  
implied emission radii of only \about0.8 km suggest they are 
quiescent LMXBs in which the x-ray emission 
is dominated by incandescence of 
the hot NS, slowly cooling from a previous accretion outburst phase. 
Using a hydrogen-atmosphere model, which modifies the opacity 
and temperature, as demonstrated with Chandra ACIS-S spectra 
for the field quiescent LMXB Cen X-4 (20), more realistic NS radii 
of \about10 km are derived (21). 
Our detailed spectral analysis of X5 and X7 also reveal 
a power law component 
(photon index 2.6 - 3) similar to that found for Cen X-4 and 
indicative of either residual accretion or a ``propeller" spin-down 
luminosity source, as well as a possible line emission  
component (modeled (21) as a collisionally excited 
Raymond-Smith plasma with kT = 1keV), indicative 
of a thin corona or perhaps wind from the system. 
The integrated fluxes for these two components are \about0.3 to 0.5 
and \about0.2 of the black body component, respectively. Although a 
statistical test (Kolmogorov-Smirnov) 
on photon arrival times of X7 shows no significant 
variations in flux, X5 shows dramatic dips and 
possible eclipses, which may also be detected in the 
optical counterpart we have identified with HST (22). Both X5 and X7 are 
most probably quiescent LMXBs given their similar 
luminosities and spectra to those 
tabulated (23) for the 6 known NS-quiescent LMXB systems 
in the field. However, the 
hot gas and possible wind component may indicate that X5 and X7 are luminous 
newly-born MSPs ablating their companions and 
still optically thick to their radio 
emission, because they are not among the currently located MSP sample (12). 
The lack of positive flux increases (i.e. flares; vs. the 
negative dips seen in X5) and otherwise 
constant flux would support, but not require, the MSP interpretation. 
 
The 13 candidate CVs (cf. Fig. 1) identified thus far 
in our HST identification program (22) have blue stellar counterparts 
indicative of accretion disks (though not all 
are yet determined to be variable).  
Apart from the quiescent LMXBs X5 and X7, all remaining ROSAT sources 
[X6, X9, X10, X11, X13, and X19; (14)] within 
our analysis region (X4 is just outside) are CV candidates. 
Spectra have been fit (21) for the brightest 8 sources (with medcts = 
135 - 2402) and most are best fit with thermal bremsstrahlung  with 
temperatures kT = 5 - 30 keV, typical of CVs (24). 
Soft black body components, as found for quiescent 
LMXBs, are generally not present. 
Three sources have unusually hard spectra: X6 is best fit 
with a thermal bremsstrahlung spectrum with kT \gsim75 keV, 
and both W8 and W15 require similarly hard spectra but 
with internal self-absorption 
column densities of NH = 1.4$\pm$0.4 and 0.7$\pm$0.2 \X 10$^{22}$ \cmsq, 
respectively. Power law fits are equally acceptable for all three 
but with photon index $\alpha$ \about1.1$\pm$0.1 
(consistent with the slope of 
a thermal bremsstrahlung spectrum). Since such a flat power law index at low 
energies ($<$6 keV) is unprecedented for 
MSPs (for which the limited spectral data (18) indicate 
$\alpha$ \about2 -- 3) or active galactic nuclei (AGN) with typical (25)
$\alpha$ \about1.7, a thermal bremsstrahlung model is more 
likely. The excess NH 
in sources W8 and W15 is evident from their ``blue" colors in Fig. 1 
(absorption of soft counts) and might suggest they are 
background AGN, probably Seyfert 2 galaxies, even though the 
number expected (from hard sources in 
Chandra deep surveys (26)) in this 
2\arcmin \X 2.5\arcmin field and exposure is \lsim 0.1.  
However, not only are their power law spectral fits 
too hard, but our identification (22) of both W8 and 
W15 with blue stars imply 
x-ray/optical flux ratios similar to values expected for CVs  
(e.g. log(\FxFv) =  +0.1 for W15). 
X-ray self-absorption is likely in high 
inclination CVs (27) and may also arise from the 
``accretion curtain" (27), or relatively cool inner disk near the 
magnetospheric-disk interface suspected to exist 
in magnetic CVs in which accretion onto the WD is channeled 
by the magnetic field. Magnetic CVs 
also typically have relatively high temperature thermal bremsstrahlung 
components, though the kT values for X6, in particular, is unusually large. 

The brightest CV candidate, X9 (medcts 
= 2402), is well fit with a thermal bremsstrahlung 
spectrum with kT = 32$\pm$6 keV 
and is coincident (0.1\arcsec) with the blue 
CV candidate V1 (28), confirming the identification by the ROSAT 
detection (14) of X9. The 70 ksec Chandra observation  
of X9 shows evidence for pulsations (false alarm probability 
\about$2\times10^{-3}$) with period P = 218.24s and modulation 
depth 30\% (Fig. 4), suggesting this is direct evidence 
for an accreting magnetic WD and thus magnetic CV. 
Although the total flux from X9 is relatively constant (Fig. 5), 
and a factor of \about2 fainter than detected with ROSAT, 
the pulsation amplitude is variable (strongest 
in the second quarter of the observation) and could instead indicate 
transient pulsations as found in some CVs (24). 
The faint CV candidate AK09 is also a possible magnetic CV. It 
was not detected with ROSAT 
and is too faint even with Chandra (medcts = 36) for spectral 
fitting but is relatively soft (red; cf. Fig. 1). The 
Chandra position is coincident (0.1\arcsec) with the blue 
variable AK09 (29) with 1.1d binary period, and originally suggested (30) as 
the counterpart of the single x-ray source detected in the original 
Einstein x-ray image of 47 Tuc (5), for which the positional 
uncertainty (3\arcsec; 90\%) included both X9 and AK09.  
HST spectra (31) now show AK09 to be 
almost identical to the magnetic CV system GK Per, with 
similarly long (2d) binary period and occasional 
outbursts. AK09 also shows 
UV outbursts (32) so that the bright Einstein source,  
which had both a thermal bremsstrahlung 
spectrum (kT = 2-6$\pm$1 keV) as well as possible 120.2s 
pulsations (30), might have been AK09 in outburst. 
Both Chandra (21) and HST (22) studies further constrain 
the magnetic CV hypothesis. 

The optically discovered CV, V2 (33), which displays dwarf nova-like 
outbursts is identified (0.1\arcsec) with 
source X19 and fit with a thermal bremsstrahlung spectrum 
with kT = 12 $\pm$6 keV, 
and the blue variable V3 (32) is the probable counterpart (0.1\arcsec) 
for X10, in contrast to the claim from ROSAT data (14). 
The Chandra spectrum of X10 is 
unique among all the (moderately) 
bright sources for which spectral fits were possible. X10 can be  
fit with a pure power law ($\alpha = 2.7 \pm$0.2). 
Its x-ray color and extreme variability, 
with dips or flares and eclipses (Fig. 5), for which a power spectrum 
reveals a significant 3.83h period, are similar to 
the quiescent LMXB candidate X5 although a black body + 
power law component (as for X5) is only acceptable if 
a Raymond-Smith component is added in absorption. 
However, its blue variable counterpart V3, with Balmer emission lines
(34), suggests accretion is ongoing. We suggest X10, and thus perhaps X5, 
may be similar to the peculiar magnetic CV AE Aqr, in which spin 
energy of the WD is partly converted to non-thermal 
radiation by a magnetic propeller, possibly 
anchored in the disk (35). Alternatively, the power law spectrum suggests 
X10 might be, as mentioned for X5, an enshrouded MSP.

The fourth group of probable source identifications 
are main sequence binaries. Of the 6 sources marked 
(Fig. 1), the brightest, identified with the 12.7h binary 
E8 (29), is the most convincing association with what may 
be chromospherically active M-S binaries. No blue excess is 
found in our HST analysis (22), and the star appears just above 
the main sequence. This is as expected for a BY Dra star, the main 
sequence analog of RS CVn systems in which x-ray emission 
is produced primarily in giant flare outbursts. Although 
the \Lx = 10$^{31.4}$ \lcgs is a factor of \about10 larger 
than the maximum found with ROSAT (36) for field BY Dra systems 
in quiescence,  
this is probably due to what appears to be a smooth decline from 
a large outburst (Fig. 5). The other 5 candidate BY Dra 
systems (Fig. 1) also show x-ray flare-like variations.  
Using our Chandra-HST astrometry, we have found at 
least 10 other optical variables without blue excesses that
are possible BY Dra systems, although without x-ray flare
variations they are less likely identifications. Few, if any, 
RS CVn systems are seen, with only 2-3 possible matches 
(\lsim0.5\arcsec) of Chandra sources with sub-giants.
\medskip

\ni
{\bf Implications for Population of Compact Objects, Binaries 
and Cluster Core}

While source types for some individual cases are uncertain 
(e.g. X10), it appears that the 108 sources near the core of 
47Tuc are a mixture of 4 source types. Their 
quasi-continuous distribution in x-ray colors (Fig. 3) is 
due to the relative insensitivity of broad-band colors to 
detailed spectral differences, particularly for 
multi-component spectra. Nevertheless, some broad differences emerge 
when the distribution of x-ray color is examined (Fig. 6). 
Most striking is that 
most of the MSPs (for which identifications are unambiguous) are 
relatively soft: 8 of the 12 with medcts \gsim5 (and thus rough 
colors available, as in Fig. 3) are softer than any of the much 
brighter CV candidates. Similarly, the two most probable quiescent LMXBs 
are softer than all of the CVs (except AK09) and are intermediate 
between the CVs and MSPs. The BY Dra candidates (BYs) are intermediate 
in color but closer to the CVs. Although the statistics are 
limited, the values for the mean, \avcolor, and standard 
deviation, \sigcolor, for each population are: 
(\avcolor, \sigcolor) = (1.8$\pm$0.3, 0.98), (-0.1$\pm$0.3, 1.09), 
(0.8$\pm$0.2, 0.56) and (1.5$\pm$0.1, 0.17) for MSPs, CVs, 
BYs, and quiescent LMXBs respectively. The corresponding distribution 
values for the unidentified (UnID) sources are 
(1.1$\pm$0.1, 1.03), or between the BYs and MSPs. The BY 
contribution to the UnIDs can be measured by a deep 
followup observation. Most of the same  UnID sources will not be detected 
if they are flare outbursts from BY Dra systems, while MSPs 
will be constant in their x-ray emission. 
Given the approximate relative numbers of the 
identified sources and the Xcolor distributions (Fig. 6), we decompose 
the 61 UnID ($>$5ct)source distribution into 2 quiescent LMXBs, 18 CVs, 11 BYs, 
and 30 MSPs. Combined with the identified sources, this 
yields approximate relative (fractional) contributions of 0.04, 0.3, 0.16 
and 0.5 for the populations of quiescent LMXBs, CVs, BYs, and MSPs with 
\Lx \gsim10$^{30}$ \lcgs,  
respectively, where the uncertainties are probably 30\%. 

We conclude the source population in 47Tuc is dominated by MSPs, 
with a total number of \gsim45-60 in the present Chandra sample (with  
\Lx \gsim 1 \X 10$^{30}$ \lcgs). These would naturally account 
for the background or unresolved red sources, suggesting 
a still larger MSP population at lower luminosities, 
if these faint sources are detected as constant flux 
sources in followup observations. Further inferences for the MSP 
population, including their spindown energy 
loss, $\dot{E}$, vs. \Lx correlations 
and radial distributions in the cluster, 
are discussed separately (19). However, the CV population may 
also extend into this faint, soft source background population 
if there are large numbers of strongly magnetic (AM Her type) 
CVs for which soft black body emission dominates. These would be 
optically very faint (like the MSPs), but would display 
marked variability (relatively long high-low states), and would thus also 
be distinguished from both BY Dra and MSP systems of comparable \Lx. 

Our estimate that the present sample contains \about30 CVs 
is \about1/3 the number predicted for 
tidal capture CV production models (37). This  may indicate 
tidal capture is relatively inefficient, but in fact the prediction of a 
large excess of low luminosity sources (\Lx \lsim10$^{30}$ \lcgs) 
may also be consistent with our ``background'' sources. 
The size of the CV sample 
is unexpected because so few dwarf novae have been found (33). 
Combined with the evidence presented here that the 
brightest CV candidates (e.g. X9) are magnetic CVs, it may be that CVs 
in globulars are indeed dominated by magnetic systems, as 
suggested from HST spectra (38), in which 
dwarf nova outbursts are suppressed, at least 
for short binary period systems. 

If only 2-3 quiescent LMXBs (X5, X7, and possibly X10) are present in 
the core with the large sample of \about100 MSPs inferred here or from 
radio counts (8), the spin-up timescales for MSPs are 
reduced to T$_{spinup}$ \about N$_{quiescent LMXBs}$/N$_{MSPs}$ \X 
T$_{MSPs}$, where T$_{MSPs}$ \about3 \X 10$^{9}$ years is the 
typical MSP spindown age. Thus T$_{spinup}$ \about1 \X 10$^8$ years, 
which is \about10\X faster than the 
\gsim$10^9$ years needed to transfer the \about0.03P$_4^{-4/3}$ \Msun
required (7) to spin up the NSs in 47Tuc to their typical \about4msec 
spin period (P$_4$ is the final period in units of 4msec) at the 
time-averaged  mass accretion rate  
\mdot \about 3 \X 10$^{-11}$\Msun yr$^{-1}$ appropriate to 
typical LMXBs in clusters with \Lx \about10$^{36.3}$ \lcgs (3) 
which may be in quiescence \about50\% of the time. Thus 
the observed quiescent LMXBs are insufficient to support the 
MSP population and alternative paths to MSP production 
may be required such as direct collapse of massive WDs in 
high \mdot CVs (7) (for which X9 may be relevant) or  
MSP formation from common envelope evolution of a NS 
with a prior generation of \about1-3\Msun binary companions 
leading to a short-lived NS-WD LMXB phase (39).  

Finally, the Chandra image allows the most sensitive mass limits for 
a black hole in the core of a globular cluster. The precise MSP 
positions and associated pulsar dispersion measures (12) imply an electron 
density (and thus ionized gas at T \about10$^4$ K) 
of n \about 0.1 cm$^{-3}$ in the 
cluster core. Bondi-Hoyle spherical accretion with radiative efficiency 
$\epsilon$  onto a compact object of 
mass M, radius R and velocity V in a gas with number density n 
and sound speed c$_s$ gives an accretion luminosity 
L$_{acc}$ = $\epsilon$ G M$^3$ m$_p$ n (V$^2$ + c$_s^2$)$^{-3/2}$
R$^{-1}$, where m$_p$ is the proton mass. 
For an isolated NS as recently considered for low luminosity 
x-ray sources in globulars (40), this gives L$_{acc}$ = 
10$^{30}$ \lcgs for 47Tuc and appropriate NS values of 
M$_{ns}$ = 1.4\Msun, R$_{ns}$ = 10 km, 
$\epsilon$ = 0.1 and V$_{ns}$ \about c$_s$ \about 10km/s. 
If the accretion energy is thermalized over the NS 
surface, the expected    
black body spectrum with kT \about15 eV is too soft  
to account for the detected Chandra sources. 
Although the thermalization process is uncertain, 
higher temperature radiation from a magnetic pole is also 
inhibited at such low accretion rates 
in the spherical case if the NS is rapidly rotating. 
However, for a black hole, with larger 
mass M$_{bh}$, emission radius \about3 
Schwarzshild radii and lower equipartition 
velocity V$_{bh}$ = V$_{ns}$(M$_{ns}$/M$_{bh}$)$^{1/2}$ 
(so that c$_s$ dominates), the 
accretion luminosity will likely arise from a much lower efficiency 
flow with $\epsilon$ \about10$^{-4}$ and 
radiate a much higher temperature spectrum (kT \about100 keV), 
such as calculated in advection (or convection) 
dominated accretion models (41) for accretion on the black 
hole x-ray source SgrA* in the galactic center.  
We estimate a soft x-ray luminosity \Lx(0.5-2.5keV) 
\about 4.5 \X 10$^{25} \epsilon_{-4}$M$_{bh}^{2}$T$_{100}$ \lcgs, 
where the uncertain radiation efficiency $\epsilon$ and 
spectral temperature T are scaled 
to 10$^{-4}$ and 100 keV, respectively. 
The Chandra upper limit of \Lx \about1 \X 10$^{31}$ \lcgs for 
the brightest source in the cluster center error circle (42) (Fig. 1) 
implies an upper limit of 
M$_{bh}$ \about470\Msun $\epsilon_{-4}^{-1/2}$T$_{100}$. 
This is consistent with the understanding 
that binary heating (2) prevents core collapse into 
a moderately massive central black hole 
and that stellar mass (\about 3-10 \Msun) 
black holes produced from the initial stars more massive than the 
NS cutoff have not coalesced in the core but rather have been 
ejected by the hardest binaries (43). A more precise cluster center 
and accretion model can improve these limits for M$_{bh}$, which 
are already below the 1700\Msun value derived from the central 
surface brightness profile (44). Similar limits for those post core 
collapse clusters which, like 47Tuc, do not contain a bright LMXB 
and for which central gas densities could be estimated from MSPs, 
would be especially interesting.

\bigskip\ni
{\bf References and Notes}

\begin{enumerate}

\item G. Meylan and C. Pryor, in {\it Dynamics of Globular Clusters}, S.
Djorgovski and G. Meylan, eds., ASP Conf. Series, {\bf 50}, 31 (1993). 

\item P. Hut et al., \pasp, {\bf 104}, 981 (1992).

\item J. E. Grindlay, {\it The Globular Cluster Galaxy Connection},  
G.H. Smith and J.P. Brodie, eds, 
{\it ASP Conf. Ser.}, {\bf 48}, 156 (1993); and 
L. Sidoli et al , \aap, 
in press (and available at http://xxx.lanl.gov/abs/astro-ph/0012383).

\item W. Lewin, J. van Paradijs and R. Taam, in {\it X-ray Binaries}, 
W. Lewin, J. van Paradijs and E. van den Heuvel, 
eds. (Cambridge Univ. Press: Cambridge, 1995), 175.

\item Precise (\lsim3\arcsec) x-ray positions for 
globular cluster LMXBs as measured by 
J.E. Grindlay, P. Hertz, J. Steiner, S. Murray, and 
A. Lightman \apj, {\bf 282}, L13 (1984) enabled mass estimates 
for the sources and subsequent optical identifications for 
several as described by L. Homer, S. Anderson, B. Margon and 
E. Deutsch, \apj, {\bf 550}, L155 (2001) and references therein.

\item G. Clark \apj, {\bf 199}, L143 (1975) and J. Katz, {\it Nature},
{\bf 253}, 698 (1975). 

\item D. Bhattacharya and E.P. van den Heuvel, {\it Physics Reports}, 
{\bf 203}, 1 (1991) provide a comprehensive review of LMXB and 
MSP evolution. The AIC model for production of MSPs directly 
from collapse of WDs was proposed to solve the possible LMXB vs. 
MSP birthrate problem in globulars by J. Grindlay and C. Bailyn, 
{\it Nature}, {\bf 336}, 48 (1988). 

\item F. Camilo, D.R. Lorimer, P. Freire, A.G. Lyne, and 
R.N. Manchester, \apj, {\bf 535}, 975 (2000).

\item The {\it Chandra X-ray Observatory} is described by 
M. Weisskopf, S. O'Dell and L. van Speybroeck, {\it Proc. SPIE}, 
{\bf 2805}, 142 (1996); the ACIS detector system is described by 
G. Garmire, J. Nousek and M. Bautz, in preparation.

\item P. Hertz and J. Grindlay, \apj {\bf 275}, 105 (1983).

\item F. Verbunt, R. Elson and J. van Paradijs, \mn, {\bf 210}, 899 
(1984)

\item P.C. Freire, F. Camilo, D.R. Lorimer, A.G. Lyne, R.N. Manchester, 
and N. D'Amico, \mn, in press (2001) (and available at 
http://xxx.lanl.gov/abs/astro-ph/0103372).

\item The core radius, \rc, is defined as the radius at which the projected 
surface brightness has decreased by a factor of 2 from its peak 
central value. It is a measure of central stellar density and can 
be related to cluster dynamical parameters (see ref. (1)). The most 
reliable determination of \rc = 23\arcsec~ for 47Tuc is given by 
J.H. Howell, P. Guhathakurta, and R.L. Gilliland, \pasp, 
{\bf 112}, 1200 (2000). 

\item F. Verbunt and G. Hasinger, \aap, {\bf 336}, 895 (1998) report 
the results of the x-ray survey of 47Tuc carried out with the 
German-US x-ray telescope ROSAT (Rontgen Satellit) and its 
High Resolution Imager (HRI) detector.

\item Wavdetect is a source detection tool employing wavelet 
smoothing to match the expected point spread function. Along 
with the spectral analysis package, Sherpa, used for spectral fits 
to the brighter (\gsim300 ct) sources, it is part of the Chandra 
Interactive Analysis Operation (CIAO) developed by the Chandra X-ray 
Center (CXC) for analysis of Chandra data and available at 
http://asc.harvard.edu/ciao/. 

\item M. Zoccali, A. Renzini, S. Ortolani, A. Bragaglia, R. Bohlin, 
E. Carretta, F. Ferraro, R. Gilmozzi, J. Holberg, G. Marconi, R. Rich, 
F. Wesemael, preprint (2001) 
(and available at http://xxx.lanl.gov/abs/astro-ph/0101485).

\item A 2D KS-test method to detect overlapping but statistics-limited 
sources has been developed by S. Metchev 
and J. Grindlay \apj, in preparation (2001); maximum likelihood 
methods are used in ref. (14).

\item W. Becker and J. Trumper, \aap, {\bf 341}, 803 (1999). 

\item J. Grindlay, F. Camilo, P. Edmonds and C. Heinke, 
in preparation (2001).

\item R.E. Rutledge, L. Bildsten, E.F. Brown, G.G. Pavlov, and 
V. E. Zavlin, \apj, in press (2001) (and available at 
http://xxx.lanl.gov/abs/astro-ph/0012400). 

\item C. Heinke, J. Grindlay and P. Edmonds, in preparation (2001). 
Spectral models fit to the cluster sources included both 
continuum (black body, bremsstrahlung and power law) and emission 
line (Raymond-Smith collisional excitation) models, as provided 
in the CIAO analysis package (see ref. 15). 

\item P. Edmonds, R. Gilliland, J. Grindlay, C. Heinke 
et al, in preparation (2001).

\item M. Garcia, J. McClintock, R. Narayan, P. Callanan and 
S. Murray, \apj, in press (2001) (and available at 
http://xxx.lanl.gov/abs/astro-ph/0012452).

\item B. Warner, {\it Cataclysmic Variables}, (Cambridge Univ. Press: 
Cambridge, 1995).

\item A. Zdjiarski in {\it ASP Conf. Series}, {\bf 161}, 16 (1999) 
provides a recent review.

\item R. Giaconni, P. Rosati, P. Tozzi, M. Nonino, G. Hasinger, 
J. Bergeron, S. Borgani, R. Gilli, R. Gilmozzi and W. Zheng, 
preprint (and available at http://xxx.lanl.gov/abs/astro-ph/0007240).

\item A. van Teesling, K. Beurmann and F. Verbunt, \aap, 
{\bf 315}, 467 (1996) suggest absorption by the disk 
in high inclination non-magnetic CVs, whereas accretion curtain 
absorption in magnetic systems is proposed by 
C. Hellier, in {\it Annapolis Workshop on Magnetic 
Cataclysmic Variables}, C. Hellier and K. Mukai, eds., 
{\it ASP Conf. Ser.}, {\bf 157}, 1 (1999).

\item F. Paresce, G. de Marchi and F. Ferraro, {\it Nature}, {\bf 360}, 
 46 (1992).

\item P. Edmonds, R. Gilliland, P.. Guhathakurta, L. Petro, A. Saha, 
and M. Shara, \apj, {\bf 468}, 241 (1996).

\item M. Auriere, L. Koch and S. Ortolani, \aap, {\bf 214}, 
113 (1989).

\item C. Knigge, M. Shara, D. Zurek, K. Long and R. Gilliland, 
preprint (and available at http://xxx.lanl.gov/abs/astro-ph/0012187).

\item D. Minniti, G. Meylan, C. Pryor, E. Phinney, B. Sams, 
and C. Tinney, \apj, {\bf 474}, L27 (1997).

\item M. Shara, L. Bergeron, R. Gilliland, A. Saha and R. Petro, 
\apj, {\bf 471}, 804 (1996). 

\item P. Edmonds et al in preparation (2001). 

\item K. Horne, in {\it Annapolis Workshop on Magnetic 
Cataclysmic Variables}, C. Hellier and K. Mukai, eds., 
{\it ASP Conf. Ser.}, {\bf 157}, 349 (1999).

\item R. Dempsey, J. Linsky, T. Fleming and J. Schmitt, \apj, 
{\bf 478}, 358 (1997). 

\item R. Di Stefano and S. Rappaport, \apj, {\bf 423}, 274 (1994).

\item  J. Grindlay, in {\it Annapolis Workshop on Magnetic 
Cataclysmic Variables}, C. Hellier and K. Mukai, eds., 
{\it ASP Conf. Ser.}, {\bf 157}, 377 (1999); and 
P. Edmonds, J. Grindlay, A. Cool, H. Cohn, P. Lugger and C. Bailyn, 
\apj, {\bf 516}, 250 (1999).

\item F. A. Rasio, E.D. Phahl and S. Rappaport, \apj, 
{\bf 532}, L47 (2000).

\item E. Pfahl and S. Rappaport, \apj, in press (2001) 
(and available at http://xxx.lanl.gov/abs/astro-ph/0009212).

\item E. Quataert and R. Narayan, \apj, {\bf 520}, 298 (1999).  

\item D. Calzetti, G. DeMarchi, F. Paresce, and M. Shara, 
\apj, {\bf 402}, L1 (1993); and G. De Marchi, F. Paresce, 
M. Stratta, R. Gilliland, R. and Bohlin, \apj, 
{\bf 468}, L51 (1996).

\item S.F. Portegies Zwart and S. McMillan, \apj, {\bf 528}, 
L17 (2000).

\item P. Guhathakurta, B. Yanny, D. Schneider and J. Bahcall, 
in {\it Dynamics of Globular Clusters}, S.
Djorgovski and G. Meylan, eds., ASP Conf. Series, {\bf 50}, 303 (1993). 

\item We thank F. Camilo, H. Cohn and R. Gilliland for helpful comments. 
This observation was conducted as part of the GTO program on the Chandra 
X-ray Observatory with support from NASA (HRC contract NAS8-38248). 
\end{enumerate}

\newpage


\small
\vspace*{-1.3in}
\hspace*{1.in}
\psfig{file=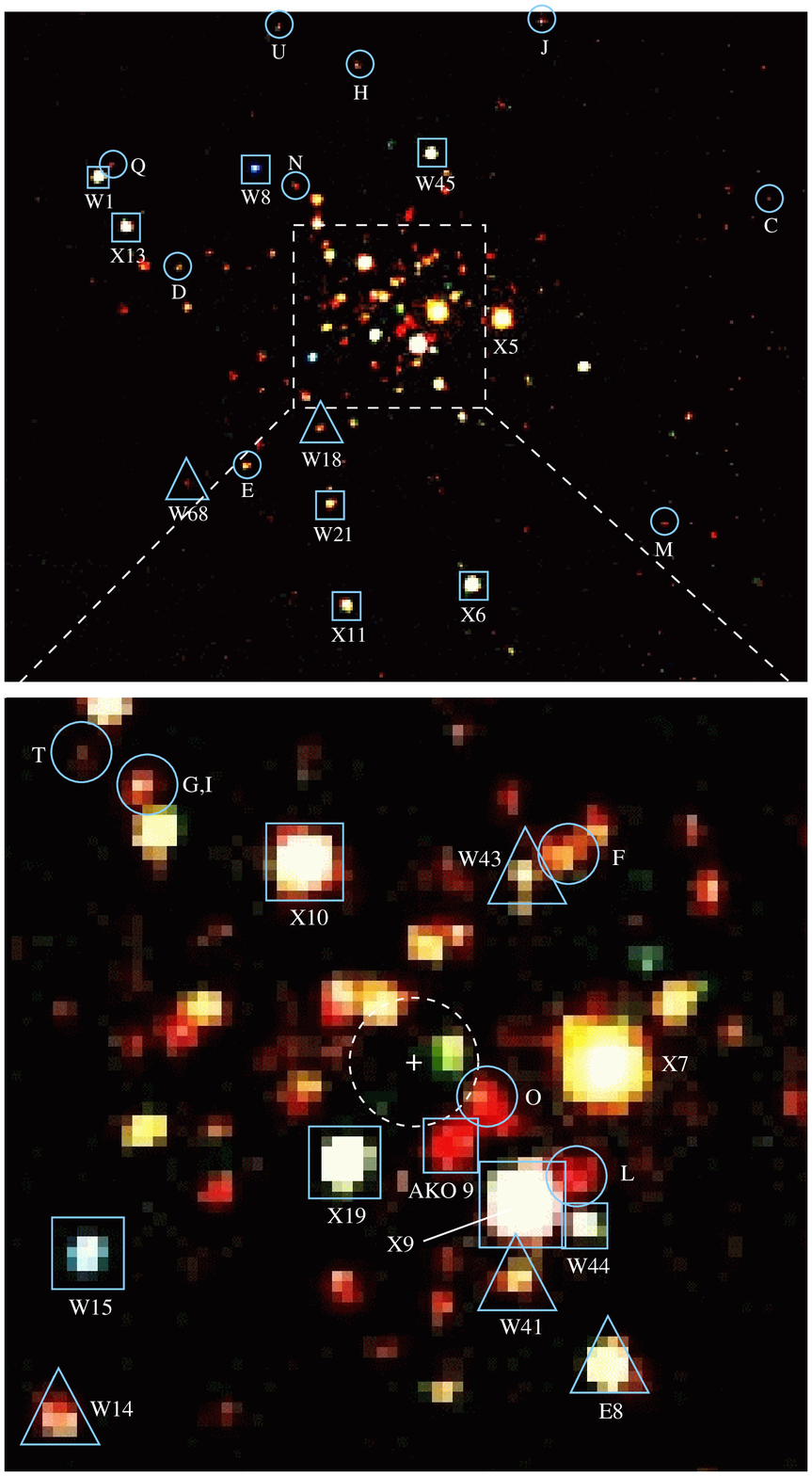,width=4.in,height=6.5in}
\medskip

\ni
{\bf Fig. 1.} {X-ray color image of the central 2\arcmin \X 2.5\arcmin
of 47Tuc. Colors are derived from counts detected in soft (0.5 - 1.2 
keV; red), intermediate (1.2 - 2 keV; green) and hard (2 - 6 keV; blue) 
bands with the I3 chip of the ACIS-I detector on Chandra. The pixel size 
(visible in enlargement of central core) is 0.4914\arcsec, and 
the enlarged central region is 35\arcsec square. Source 
identifications are: MSPs (circles), marked with source letter names 
(12); quiescent LMXBs X5 and X7; CV candidates (squares), marked 
with ROSAT Xnumbers (14) 
or other Chandra Wnumbers; and possible flaring BY Dra systems, 
or M-S binaries (triangles), marked 
with E8 (29) or Chandra Wavdetect source numbers 
(Wnumbers). Precise positions, fluxes, hardness 
ratios, bright source spectra and variability are tabulated for all 108 
sources shown here (73 unidentified are not labeled, for clarity) 
in our full-field analysis (21), with only key 
values given here in the text. The mean cluster center (41) and 
3$\sigma$  error (3\arcsec~ radius dashed circle) are marked; other 
symbols (squares, etc.) are centered on sources but with sizes 
much greater than positional uncertainties.}
 
\newpage

\vspace*{-1.in}

\hspace*{0.1in}
\psfig{file=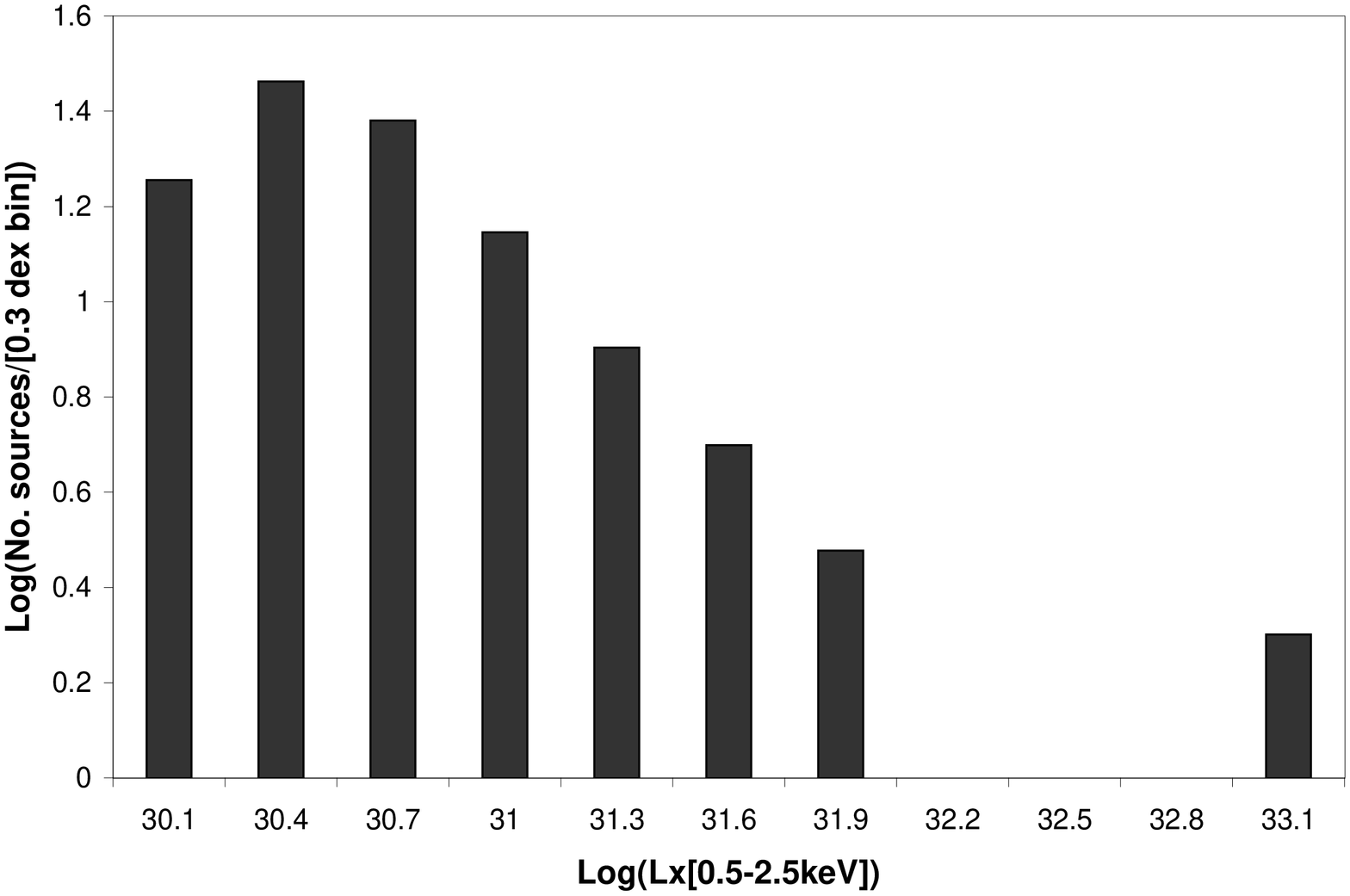,width=6.in,height=4.in,angle=-90.}

\vspace*{-1.3in}
\hspace*{1.2in}
\psfig{file=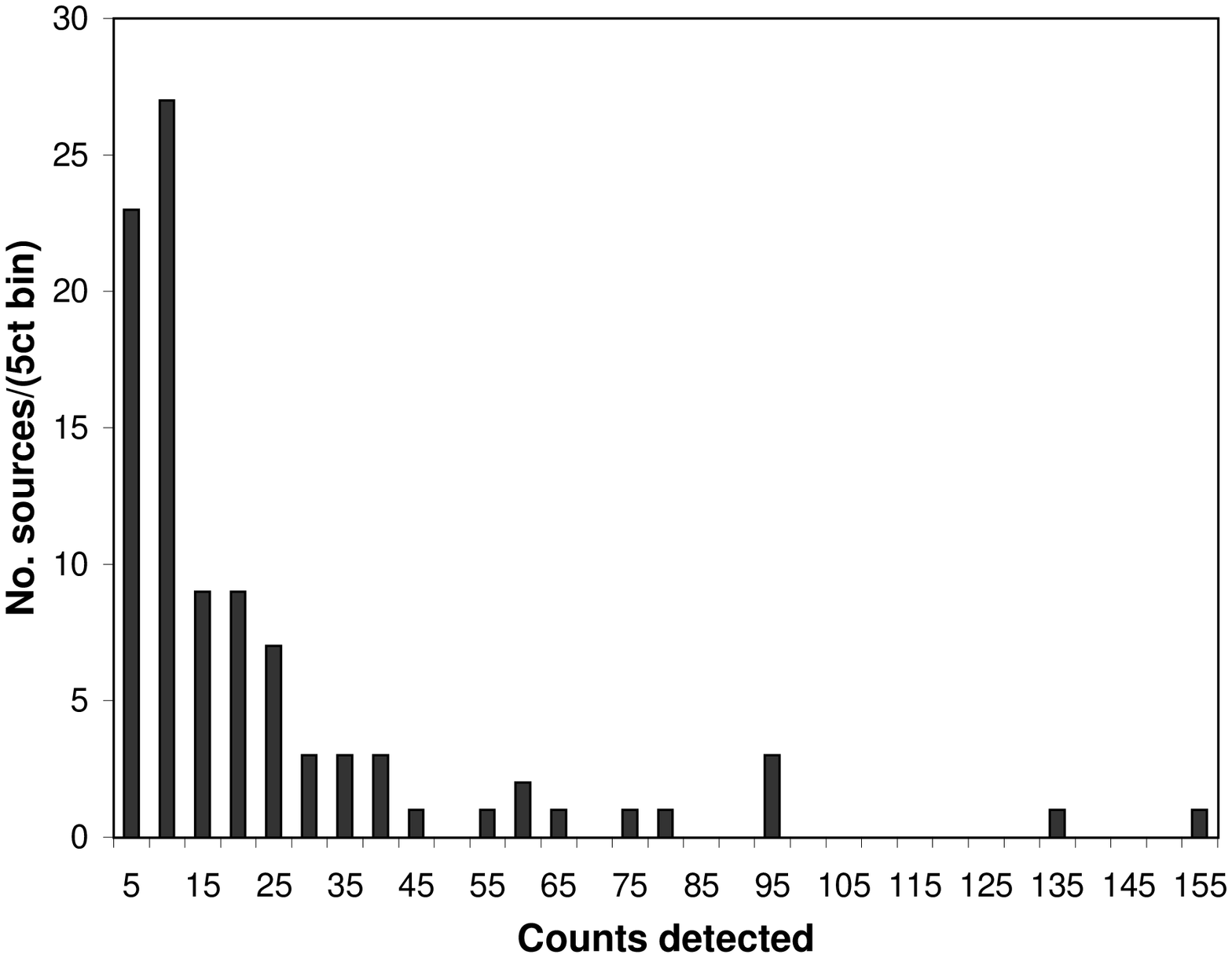,width=5.in,height=4.in,angle=-90.}

\vspace*{-0.6in}
 
\ni
{\bf Fig. 2.} X-ray luminosity and counts distributions of  
sources detected with $>$2.5cts in central 
2\arcmin \X 2.5\arcmin of 47Tuc. Counts detected in 
medium band (0.5 - 4.5 keV) are corrected for instrument and 
telescope response, 5.0 kpc cluster distance (16) and  
known interstellar absorption of A$_V$ = 0.12 (or equivalent 
column density NH = 2.4 \X 10$^{20}$ \cmsq) and an assumed 
thermal bremsstrahlung spectrum with kT=1 keV to give source 
luminosity in the 0.5-2.5 keV band for direct comparison with 
ROSAT results (14). The extended energy band of Chandra 
(0.5 - 8 keV) is utilized for hardness ratio and spectral analysis, which 
also enable conversion of \Lx between the Chandra and ROSAT bands.
{\bf (A)} log source counts vs. log \Lx, with log \Lx value given 
at the center of 0.3dex bins, {\bf (B)} 
linear source counts vs. actual counts detected , with value 
given at center of 5ct wide bins. Corresponding log \Lx values range 
from 29.9 to 31.4. The 96 sources plotted in (B) account for only 13\% 
of the total detected x-ray flux from the cluster 
central region, whereas the brightest two sources (probable 
quiescent LMXBs X5 and X7) contribute 57\% of the total counts. 

\newpage

\psfig{file=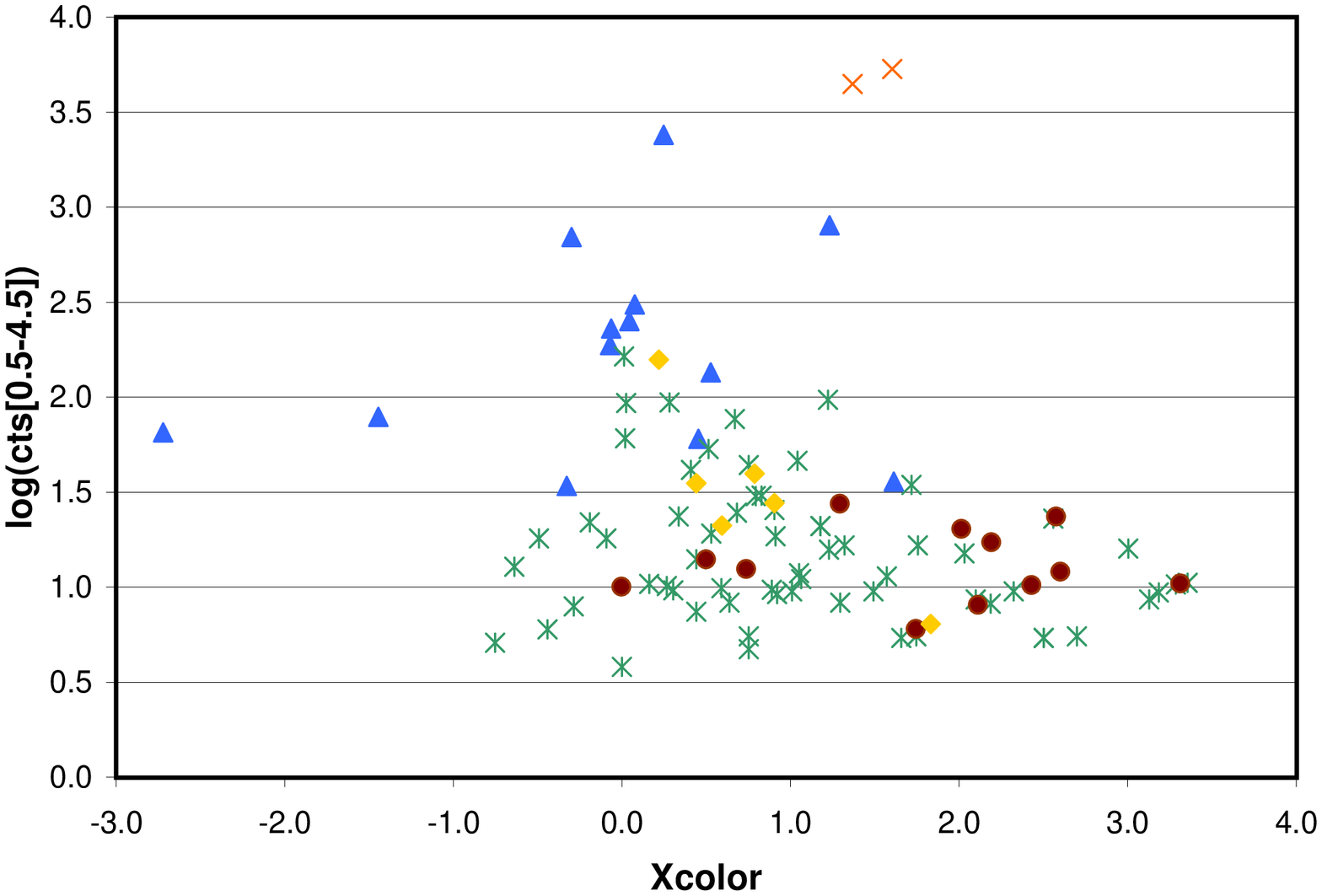,width=7.in,height=5.5in,angle=-90.}
\bigskip

\ni
{\bf Fig. 3.} X-ray brightness (medcts= 0.5-4.5keV) vs. color 
(ratio of soft(0.5-1.5keV)/hard(1.5-6keV) counts)
distribution for sources with medcts \gsim5 cts. 
Source types are labeled for the sources identified with 12 MSPs 
(red $\bullet$), 2 quiescent LMXBs (orange X), 13 CVs 
(blue $\triangle$) and 6 M-S binaries (BY Dra systems)  
in flare outbursts (yellow $\Diamond$). The 61 remaining 
(green $\ast$) are unidentified. Source hardness increases to left. 

\newpage

\hspace*{-0.5in}
\psfig{file=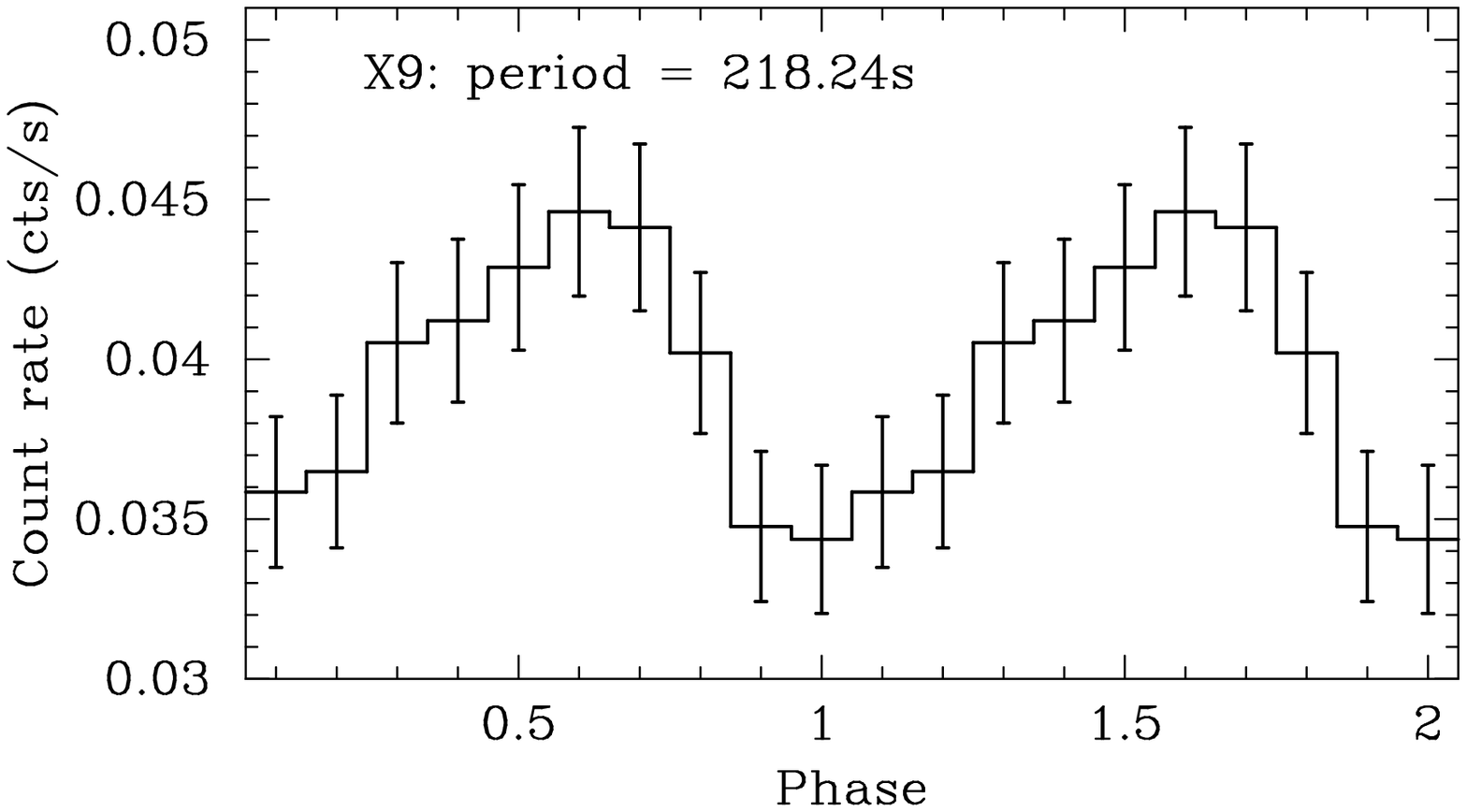,width=6.7in,height=3.7in}
\bigskip

\ni
{\bf Fig. 4.} Folded light curve for the brightest CV candidate, X9, 
showing its probable detection as a 218.24s pulsed source and thus 
identification as an accreting magnetic WD, or magnetic CV system (a NS 
x-ray pulsar with such a long period would be expected only in 
a wind-fed high mass x-ray binary found only near regions of 
massive star formation in the Galaxy, not in globular clusters).

\newpage

\vspace*{-1.in}

\hspace*{0.5in}
\psfig{file=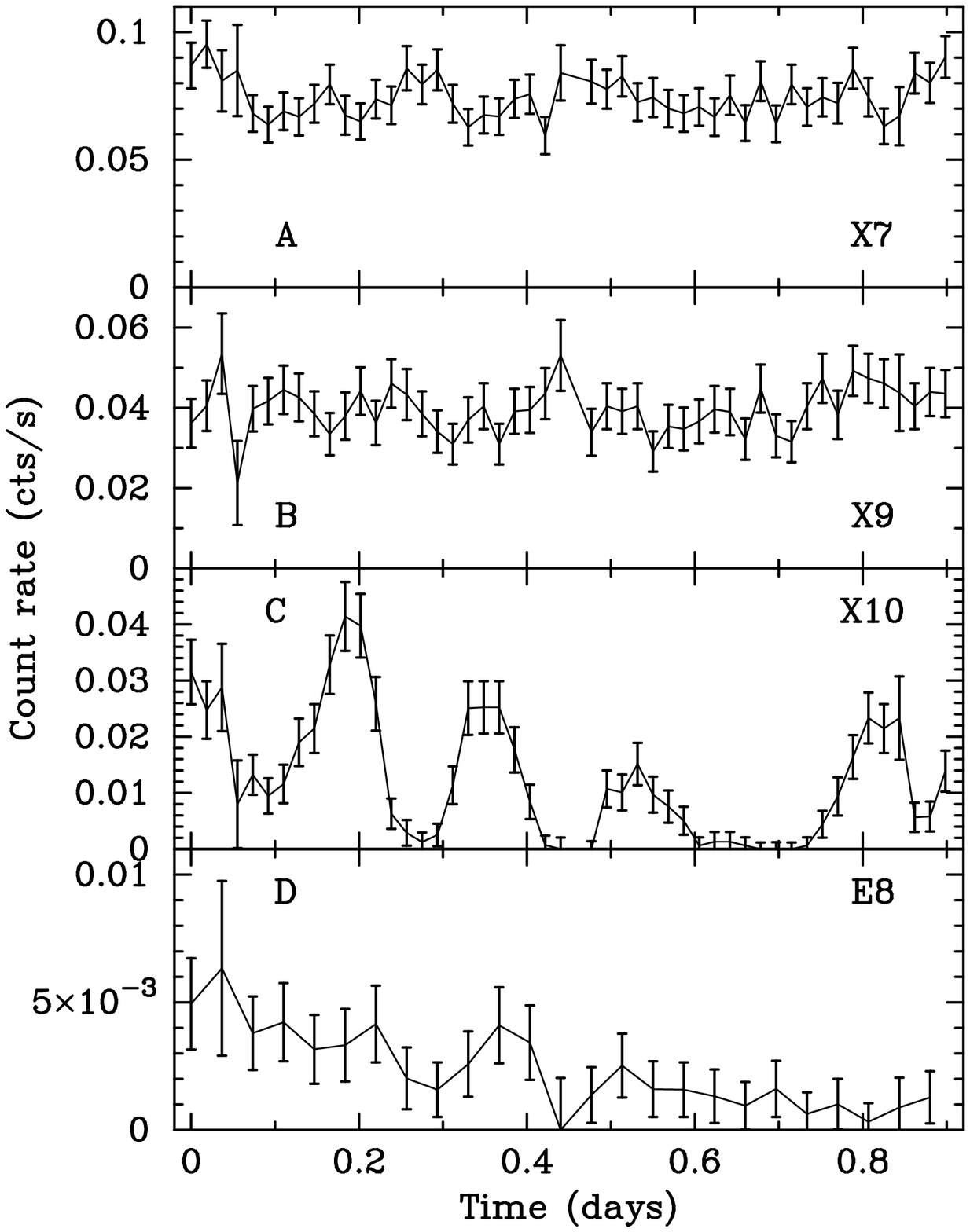,width=5.5in,height=7.in}

\ni
{\bf Fig. 5.} Light curves, over the full 74ks (duration) observation,
for 3 of the 4 source types (all MSP fluxes are constant, 
within statistics). {\bf (A)} quiescent LMXB source X7;  
{\bf (B)} CV candidates  
X9 (cf. Fig. 4), showing flickering, and {\bf (C)} X10, showing dips and 
eclipses with 3.8h period (similar to X5); and  {\bf (D)} M-S binary 
candidate E8, showing smooth decline from probable large flare.

\newpage

\psfig{file=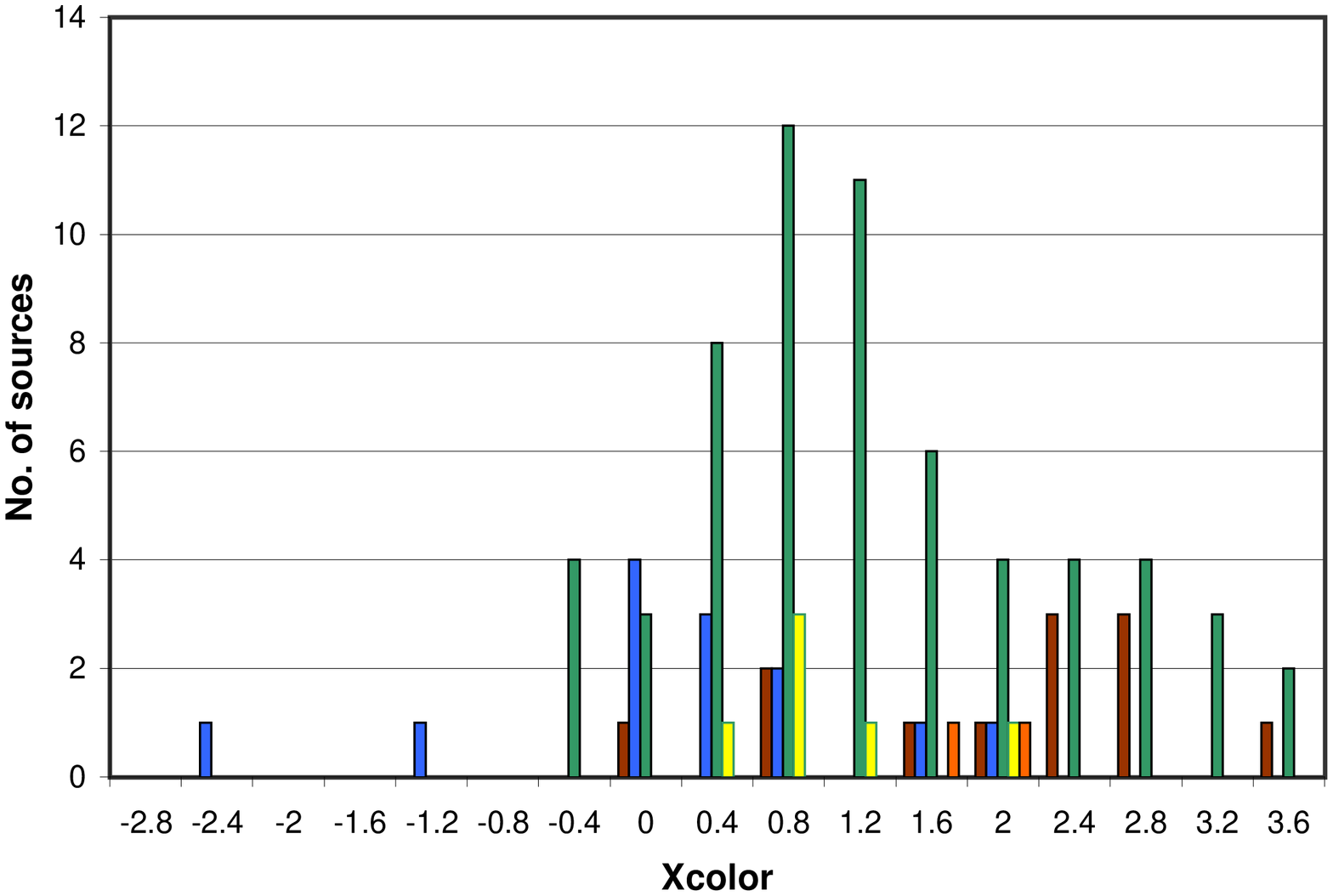,width=7.in,height=5.in,angle=-90.}

\bigskip

\ni
{\bf Fig. 6.} Distributions of x-ray color (cf. Fig. 3) for 94 
sources with medcts \gsim5 counts and proposed source identifications 
(see Figs. 1 and 3; source types are color coded same as 
in Fig. 3). M-S binaries are BY Dra systems in 
flaring outbursts. Means and standard deviations for each 
source distribution are given in the text.

\end{document}


\hoffset=-0.5truein

\begin{center}

\begin{deluxetable}{ccccccccc}
\tablewidth{7.5truein}
\tablecaption{47 Tuc Core X-Ray Sources (Grindlay et al 2001, {\it Science} On Line, 17 May 2001)}
 
\tablehead{
\colhead{Src$^{a}$} & \colhead{RA\thinspace$^{b}$} & \colhead{Dec\thinspace$^{b}$} &
\colhead{Medcts\thinspace$^{c}$} & \colhead{Sfcts} & \colhead{Hdcts} & 
\colhead{Sf/Hd\thinspace$^{d}$} &
\colhead{$L_{X}$\thinspace$^{e}$} & \colhead{ID\thinspace$^{f}$}

}
\startdata

W1 & 0:24:16.969$\pm$0.007 &   -72:04:27.18$\pm$0.03 & 229.7 & 117.4 & 124.6 & 0.9 & 4.8e31 &  CV \\
W2 & 0:24:15.89$\pm$0.007 &   -72:04:36.37$\pm$0.03 & 187.3 &  98.1 & 104.6 & 0.9 & 3.9e31 &  X13,CV \\ 
W3 & 0:24:15.163$\pm$0.02 &   -72:04:43.36$\pm$0.09 &  34.6 &  29.7 &   6.1 & 4.8 & 7.3e30 &  - \\ 
W4 & 0:24:13.464$\pm$0.01 &   -72:04:51.13$\pm$0.08 &  30.3 &  22.1 &  10.3 & 2.1 & 6.4e30 &  - \\ 
W5 & 0:24:12.58$\pm$0.02 &   -72:04:41.10$\pm$0.05 &   8.2 &   7.5 &     1 & 7.7 & 1.7e30 &  - \\ 
W6 & 0:24:11.649$\pm$0.03 &   -72:05:03.69$\pm$0.1 &   9.4 &   9.4 &     0.5 &   20  & 2.0e30 &  - \\ 
W7 & 0:24:11.125$\pm$0.02 &   -72:05:20.14$\pm$0.06 &  23.4 &  21.5 &     2 & 11 & 4.9e30 &  MSP-E \\ 
W8 & 0:24:10.763$\pm$0.01 &   -72:04:25.71$\pm$0.05 &  65.2 &   6.4 &  78.5 & 0.1 & 1.4e31 &  CV \\
W9 & 0:24:10.512$\pm$0.02 &   -72:05:06.35$\pm$0.2 &   9.6 &   5.7 &   4.3 & 1.3 & 2.0e30 &  - \\
W10 & 0:24:10.07$\pm$0.02 &   -72:04:41.23$\pm$0.09 &  18.5 &  13.2 &   5.7 & 2.3 & 3.9e30 &  - \\
W11 & 0:24:09.844$\pm$0.02 &   -72:03:59.69$\pm$0.1 &  12.4 &   7.7 &   3.9 & 2 & 2.6e30 &  MSP-U \\
W12 & 0:24:09.514$\pm$0.02 &   -72:05:04.58$\pm$0.1 &   8.3 &   6.6 &     2 &  3.3 & 1.7e30 &  - \\
W13 & 0:24:09.193$\pm$0.03 &   -72:04:28.83$\pm$0.1 &  10.2 &   9.4 &     1 & 9.1 & 2.1e30 &  MSP-N \\
W14 & 0:24:08.78$\pm$0.02 &   -72:05:07.46$\pm$0.08 &  39.5 &  26.6 &  12.9 & 2.1 & 8.3e30 &  M-S \\
W15 & 0:24:08.496$\pm$0.01 &   -72:05:00.23$\pm$0.05 &  78.8 &  18.5 &  69.9 & 0.3 & 1.7e31 &  CV \\
W16 & 0:24:08.299$\pm$0.01 &   -72:04:35.68$\pm$0.06 &  93.7 &  55.3 &  42.6 & 1.3 & 2.0e31 &  - \\
W17 & 0:24:08.32$\pm$0.009 &   -72:04:31.39$\pm$0.04 &  96.6 &  74.5 &  24.2 & 3.1 & 2.0e31 &  - \\
W18 & 0:24:08.17$\pm$0.02 &   -72:05:13.29$\pm$0.09 &  21.1 &  12.8 &   7.4 & 1.7 & 4.4e30 &  M-S \\
W19 & 0:24:07.97$\pm$0.02 &   -72:04:39.64$\pm$0.07 &  17.1 &  15.1 &     2 & 7.7 & 3.6e30 &  MSP-G,I \\
W20 & 0:24:07.924$\pm$0.02 &   -72:04:54.75$\pm$0.05 &  53.4 &  33.9 &  21.1 & 1.6 & 1.1e31 &  - \\
W21 & 0:24:07.773$\pm$0.01 &   -72:05:27.10$\pm$0.05 &  60.3 &  37.2 &  24.5 & 1.5 & 1.3e31 &  CV \\
W22 & 0:24:07.832$\pm$0.02 &   -72:05:24.38$\pm$0.09 &  10.4 &   5.8 &     5 & 1.2 & 2.2e30 &  - \\
W23 & 0:24:07.804$\pm$0.009 &   -72:04:41.57$\pm$0.05 &  93.2 &  48.8 &  47.7 & 1 & 2.0e31 &  - \\
W24 & 0:24:07.376$\pm$0.02 &   -72:04:49.41$\pm$0.05 &  41.5 &  40.1 &    12 &  3.3 & 8.7e30 &  - \\
W25 & 0:24:07.141$\pm$0.009 &   -72:05:45.67$\pm$0.04 & 134.7 &  86.5 &  53.3 & 1.6 & 2.8e31 &  X11,CV \\
W26 & 0:24:06.869$\pm$0.01 &   -72:05:12.29$\pm$0.08 &  25.5 &  17.7 &   7.7 & 2.3 & 5.4e30 &  - \\
W27 & 0:24:06.397$\pm$0.003 &   -72:04:42.97$\pm$0.01 & 802.4 & 626.5 & 201.7 & 3.1 & 1.7e32 &  X10,V3,CV \\
W28 & 0:24:06.039$\pm$0.02 &   -72:05:01.6$\pm$0.1 &    16 &  15.9 &     1 & 16.7 & 3.4e30 &  - \\
W29 & 0:24:06.064$\pm$0.01 &   -72:04:48.93$\pm$0.07 &  41.3 &  27.7 &    19 & 1.4 & 8.7e30 &  - \\
W30 & 0:24:06.013$\pm$0.005 &   -72:04:56.20$\pm$0.03 & 307.3 & 166.6 & 155.5 & 1.1 & 6.5e31 &  X19,V2,CV \\
W31 & 0:24:05.662$\pm$0.03 &   -72:05:04.51$\pm$0.08 &   5.5 &     6 &     0.5 &    12 & 1.2e30 &  - \\
W32 & 0:24:05.656$\pm$0.01 &   -72:04:49.15$\pm$0.04 &  76.8 &  50.6 &  27.3 & 1.9 & 1.6e31 &  - \\
W33 & 0:24:05.41$\pm$0.02 &   -72:04:21.36$\pm$0.1 &  18.1 &   7.7 &  12.1 & 0.6 & 3.8e30 &  - \\
W34 & 0:24:05.216$\pm$0.01 &   -72:04:46.59$\pm$0.07 &  30.2 &    22 &  10.6 & 2.1 & 6.3e30 &  - \\
W35 & 0:24:05.02$\pm$0.02 &   -72:05:06.22$\pm$0.1 &  23.5 &  13.9 &  10.2 & 1.4 & 4.9e30 &  - \\
W36 & 0:24:04.967$\pm$0.02 &   -72:04: 55.4$\pm$0.08 &  35.8 &  30.9 &     7 & 4.3 & 7.5e30 &  AKO9,CV \\
W37 & 0:24:05.007$\pm$0.02 &   -72:04:51.27$\pm$0.09 &  35.2 &  21.8 &  14.5 & 1.5 & 7.4e30 &  - \\
W38 & 0:24:04.902$\pm$0.01 &   -72:04:46.19$\pm$0.07 &    14 &     9 &     6 & 1.5 & 2.9e30 &  - \\
W39 & 0:24:04.637$\pm$0.02 &   -72:04:53.71$\pm$0.1 &  20.2 &  19.2 &     3 & 6.3 & 4.2e30 &  MSP-O \\
W40 & 0:24:04.62$\pm$0.02 &   -72:04: 34.1$\pm$0.2 &  22.9 &  21.1 &     2 & 11.1 & 4.8e30 &  - \\
W41 & 0:24:04.345$\pm$0.02 &   -72:05:01.35$\pm$0.08 &  27.6 &  20.7 &     9 & 2.3 & 5.8e30 &  M-S \\
W42 & 0:24:04.263$\pm$0.002 &   -72:04:57.97$\pm$0.01 &  2402 & 1451.4 & 1154.8 & 1.3 & 5e32 &  X9,V1,CV \\
W43 & 0:24:04.276$\pm$0.01 &   -72:04:43.52$\pm$0.1 &  21.9 &    11 &  13.1 & 0.8 & 4.6e30 &  M-S \\
W44 & 0:24:03.703$\pm$0.01 &   -72:04:59.03$\pm$0.05 &  24.5 &    18 &    23 & 0.8 & 5.1e30 &  CV \\
W45 & 0:24:03.776$\pm$0.006 &   -72:04:22.96$\pm$0.02 & 250.6 & 134.1 & 128.9 & 1 & 5.3e31 &  CV \\
W46 & 0:24:03.515$\pm$0.001 &   -72:04:51.91$\pm$0.01 & 5332.2 & 4329.5 & 989.7 & 4.3 & 1.1e33 &  X7,qLMXB \\
W47 & 0:24:03.464$\pm$0.008 &   -72:05:05.20$\pm$0.03 &   157 &  88.5 &  72.3 & 1.2 & 3.3e31 &  E8,M-S \\
W48 & 0:24:03.217$\pm$0.02 &   -72:04:29.62$\pm$0.1 &  15.7 &  12.4 &     4 & 3.1 & 3.3e30 &  - \\
W49 & 0:24:03.083$\pm$0.02 &   -72:04:47.28$\pm$0.07 &  12.8 &     5 &     9 &  0.6 & 2.7e30 &  - \\
W50 & 0:24:03.14$\pm$0.02 &   -72:04:26.54$\pm$0.08 &  16.6 &  13.5 &     4 &  3.3 & 3.5e30 &  - \\
W51 & 0:24:02.806$\pm$0.01 &   -72:04:49.04$\pm$0.06 &  46.3 &  33.9 &    13 & 2.6 & 9.7e30 &  - \\
W52 & 0:24:02.597$\pm$0.02 &   -72:05:38.41$\pm$0.1 &  11.8 &   7.9 &     3 & 2.6 & 2.5e30 &  - \\
W53 & 0:24:02.533$\pm$0.01 &   -72:05:11.25$\pm$0.06 &  60.6 &  32.9 &  32.3 & 1 & 1.3e31 &  - \\
W54 & 0:24:02.522$\pm$0.02 &   -72:04:44.24$\pm$0.09 &  20.9 &  15.7 &   5.3 & 2.9 & 4.4e30 &  - \\
W55 & 0:24:02.226$\pm$0.03 &   -72:04:50.63$\pm$0.1 &   9.9 &   6.9 &     4 & 1.7 & 2.1e30 &  - \\
W56 & 0:24:02.141$\pm$0.004 &   -72:05:41.94$\pm$0.02 &   695 & 313.8 & 412.9 & 0.8 & 1.5e32 &  X6,CV \\
W57 & 0:24:01.37$\pm$0.03 &   -72:04:45.55$\pm$0.07 &  10.1 &     6 &   4.7 & 1.3 & 2.1e30 &  - \\
W58 & 0:24:00.968$\pm$0.001 &   -72:04:53.18$\pm$0.01 & 4435.2 & 3455.7 & 981.4 & 3.6 & 9.2e32 &  X5,qLMXB \\
W59 & 0:24:00.627$\pm$0.02 &   -72:05:54.13$\pm$0.1 &    18 &   9.1 &   9.9 & 0.9 & 3.8e30 &  - \\
W60 & 0:24:00.043$\pm$0.02 &   -72:05:12.85$\pm$0.09 &   5.4 &   4.6 &     1 & 4.5 & 1.1e30 &  - \\
W61 & 0:23:59.877$\pm$0.02 &   -72:05:04.65$\pm$0.1 &  20.9 &  14.8 &     5 & 2.9 & 4.4e30 &  - \\
W62 & 0:23:59.434$\pm$0.02 &   -72:04:47.79$\pm$0.2 &   9.5 &   7.9 &     2 & 4 & 2.0e30 &  - \\
W63 & 0:23:59.392$\pm$0.03 &   -72:03:58.81$\pm$0.08 &    10 &   5.9 &   5.9 &    1 & 3.0e30 &  MSP-J \\
W64 & 0:23:57.699$\pm$0.007 &   -72:05:01.93$\pm$0.03 & 164.2 &  88.4 &  87.4 & 1 & 3.4e31 &  - \\
W65 & 0:23:53.548$\pm$0.02 &   -72:05:11.20$\pm$0.1 &   9.7 &   6.8 &     3 & 2.3 & 4.0e30 &  - \\
W66 & 0:24:16.023$\pm$0.03 &   -72:04:51.56$\pm$0.2 &  10.3 &  10.3 &    0.5 &  20 & 2.2e30 &  - \\
W67 & 0:24:13.818$\pm$0.03 &   -72:04:43.83$\pm$0.1 &  13.9 &   8.9 &   5.6 & 1.6 & 2.9e30 &  MSP-D \\
W68 & 0:24:13.477$\pm$0.02 &   -72:05:23.42$\pm$0.2 &   6.4 &   5.4 &     1 & 5.3 & 1.3e30 &  M-S \\
W69 & 0:24:12.643$\pm$0.02 &   -72:04:22.38$\pm$0.2 &   3.9 &   2.8 &     2 & 1.4 & 8.2e29 &  - \\
W70 & 0:24:11.915$\pm$0.02 &   -72:04:43.84$\pm$0.1 &    11 &     8 &     3 & 2.6 & 2.3e30 &  - \\
W71 & 0:24:10.642$\pm$0.01 &   -72:05:16.37$\pm$0.2 &   8.6 &   6.9 &     1 & 7.1 & 1.8e30 &  - \\
W72 & 0:24:10.484$\pm$0.02 &   -72:04:59.92$\pm$0.1 &  16.6 &  15.1 &     3 &  5 & 3.5e30 &  - \\
W73 & 0:24:07.224$\pm$0.02 &   -72:04:57.39$\pm$0.1 &  13.5 &   8.1 &     2 & 4 & 2.8e30 &  - \\
W74 & 0:24:06.686$\pm$0.02 &   -72:04:06.74$\pm$0.1 &  10.4 &  10.6 &     0.5 & 21 & 2.2e30 &  MSP-H \\
W75 & 0:24:06.429$\pm$0.02 &   -72:04:52.79$\pm$0.1 &  24.6 &    15 &     8 & 2 & 5.2e30 &  - \\
W76 & 0:24:6.436$\pm$0.04 &   -72:04:30.11$\pm$0.1 &   5.1 &     2 &     4 &  0.5 & 1.1e30 &  - \\
W77 & 0:24:03.92$\pm$0.02 &   -72:04:42.71$\pm$0.08 &  27.3 &  19.8 &     6 &  3.3 & 5.7e30 &  MSP-F,S \\
W78 & 0:24:03.516$\pm$0.02 &   -72:05:22.54$\pm$0.1 &   4.2 &   4.3 &     0.5 & 8.6 & 8.8e29 &  - \\
W79 & 0:24:03.565$\pm$0.02 &   -72:04:41.50$\pm$0.1 &  19.1 &    14 &   8.6 & 1.6 & 4.0e30 &  - \\
W80 & 0:24:02.658$\pm$0.02 &   -72:04:41.47$\pm$0.1 &  11.4 &   8.5 &     2 & 4.2 & 2.4e30 &  - \\
W81 & 0:24:01.775$\pm$0.03 &   -72:05:12.30$\pm$0.1 &   7.9 &     4 &   5.2 &  0.8 & 1.7e30 &  - \\
W82 & 0:24:01.434$\pm$0.02 &   -72:04:41.72$\pm$0.1 &   8.6 &   8.9 &     0.5 &  18 & 1.8e30 &  - \\
W83 & 0:24:00.977$\pm$0.03 &   -72:04:13.70$\pm$0.1 &   9.5 &   7.6 &     3 & 2.6 & 2.0e30 &  - \\
W84 & 0:24:00.424$\pm$0.04 &   -72:05:46.55$\pm$0.2 &     6 &     2 &     3 &  0.7 & 1.3e30 &  - \\
W85 & 0:23:59.404$\pm$0.04 &   -72:04:38.46$\pm$0.1 &   9.5 &   8.5 &     1 & 8.3 & 2.0e30 &  - \\
W86 & 0:23:58.988$\pm$0.02 &   -72:05:09.70$\pm$0.1 &   3.8 &     3 &     1 & 3 & 8e29 &  - \\
W87 & 0:23:57.884$\pm$0.04 &   -72:04:52.76$\pm$0.08 &   7.4 &   4.5 &     3 & 1.5 & 1.6e30 &  - \\
W88 & 0:23:54.175$\pm$0.03 &   -72:05:14.50$\pm$0.1 &   4.6 &     3 &     2 & 1.5 & 1.7e30 &  - \\
W89 & 0:23:52.544$\pm$0.03 &   -72:05:32.99$\pm$0.1 &     3 &     3 &     0.5 & 6 & 1.6e30 &  - \\
W90 & 0:23:50.806$\pm$0.03 &   -72:05:41.51$\pm$0.1 &   8.3 &   5.4 &     3 & 1.8 & 2.1e30 &  - \\
W91 & 0:24:17.55$\pm$0.05 &   -72:03:58.28$\pm$0.1 &   3.7 &     0.5 &     4 &  0.13 & 7.8e29 &  - \\
W92 & 0:24:15.678$\pm$0.02 &   -72:04:41.31$\pm$0.2 &   9.2 &     7 &     3 & 2.3 & 1.9e30 &  - \\
W93 & 0:24:12.017$\pm$0.02 &   -72:05:08.06$\pm$0.2 &   3.8 &     2 &     2 &    1 & 8.0e29 &  - \\
W94 & 0:24:11.78$\pm$0.03 &   -72:05:14.61$\pm$0.09 &   5.5 &     4 &     2 &  2 & 1.2e30 &  - \\
W95 & 0:24:10.861$\pm$  0 &   -72:04:15.84$\pm$0.1 &   2.9 &     3 &     0.5 &  6 & 6.1e29 &  - \\
W96 & 0:24:07.281$\pm$0.02 &   -72:05:19.58$\pm$0.2 &   4.4 &     4 &     1 & 4 & 9.2e29 &  - \\
W97 & 0:24:05.379$\pm$0.02 &   -72:05:11.12$\pm$0.1 &   5.5 &     5 &     1 &  5 & 1.2e30 &  - \\
W98 & 0:24:05.056$\pm$0.03 &   -72:05:02.77$\pm$0.1 &  10.5 &    11 &     0.5 &  22 & 2.2e30 &  - \\
W99 & 0:24:02.37$\pm$0.04 &   -72:05:50.66$\pm$0.09 &   4.7 &     4 &     2 &  2 & 9.9e29 &  - \\
W100 & 0:24:01.368$\pm$0.02 &   -72:04:27.26$\pm$0.2 &   5.4 &     5 &     0.5 &    10 & 1.1e30 &  - \\
W101 & 0:23:59.974$\pm$0.02 &   -72:05:31.93$\pm$0.2 &   5.4 &     5 &     0.5 &    10 & 1.1e30 &  - \\
W102 & 0:23:57.687$\pm$0.05 &   -72:05:42.98$\pm$0.1 &   2.8 &     3 &     0.5 &    6 & 5.9e29 &  - \\
W103 & 0:23:51.996$\pm$0.04 &   -72:04:29.31$\pm$0.1 &   1.9 &     2 &   0.5 &   4 & 4.0e29 &  - \\ 
W104 & 0:24:16.49 & -72:04:25.15 &  8 &  7 &    1 & 7 &   1.7e30 &
MSP-Q \\
W105 & 0:24:08.54   &   -72:04:38.91 &  6   &   5   &   1 & 5  & 1.3e30 &
MSP-T \\
W106 & 0:24:03.77  & -72:04:56.90 &  12  &   11 &    1 & 11 & 2.5e30
& MSP-L \\ 
W107 & 0:23:54.49  &       -72:05:30.72  &  3 &    3  &   0.5 & 6
& 2.3e30 & MSP-M \\
W108 & 0:23:50.35  &       -72:04:31.49  & 1 &   1  &   0.5 & 2   & 1.3e30 & MSP-C \\

\enddata
\tablecomments{$^{a}$\thinspace Chandra source number. 
$^{b}$\thinspace RA (right ascension) and Dec 
(declination) coordinates are given as epoch J2000. 
Errors are formal $1\sigma$
centroiding errors from Chandra x-ray detection routine wavdetect. 
The final five MSPs were not formally detected by
wavdetect, so the radio positions  of Freire et al, 2001 (12) are given 
without errors. $^{c}$\thinspace Medcts=counts in 0.5-4.5 keV band,
Sfcts=counts in 0.5-1.5 keV band, Hdcts=counts in 1.5-6 keV band.
Values of 0.5 counts are estimated upper limits.  $^{d}$\thinspace
Softness ratio, defined as Sfcts/Hdcts [in Figs. 3 and 6, 
Xcolor = 2.5log(Sfcts/Hdcts)].  $^{e}$\thinspace
Intrinsic luminosities in ROSAT 
(0.5-2.5 keV) band, with multiplicative exposure corrections (=1.4, 2.0, 1.7, 2.6, 
1.2, 2.5 and 1.9 for sources W63(MSPJ), W65, W88, W89, W90, MSPC and 
MSPM, respectively), based on 1 keV thermal
bremsstrahlung spectrum, 
$N_H=2.4\times10^{20}$ cm${^-2}$, d=5.0 kpc.  $^{f}$\thinspace X
number from ROSAT (14) observations (e.g. X5), certain
optical or radio counterparts (e.g. V1, AKO9, MSP-D), 
and class identifications: 
qLMXB=quiescent LMXB, M-S=main sequence binary, CV=cataclysmic
variable. Detailed discussion of of individual x-ray source spectra 
and variability is deferred to Heinke et al. 2001 (21) and 
probable optical counterparts are discussed by 
Edmonds et al. 2001 (22).}

\end{deluxetable}

\end{center}